\documentclass[aipproc]{revtex4}

\usepackage{amssymb}
\usepackage[pdftex]{graphicx}
\usepackage[cmex10]{amsmath}
\usepackage{array}
\usepackage[tight,footnotesize]{subfigure}

\usepackage{longtable}
\usepackage{afterpage}
\usepackage{float}
\usepackage{lscape}
\usepackage{psfrag}

\usepackage[toc,page]{appendix}

\usepackage{lastpage}
\usepackage{fancyhdr}
\include{commands}

\usepackage{indentfirst}

\begin{document}

\title{New Superconducting Toroidal Magnet System for IAXO, \\the International AXion Observatory}

\author{I. Shilon}
\email{idan.shilon@cern.ch}
\author{A. Dudarev}
\author{H. Silva}
\author{U. Wagner}
\author{H. H. J. ten Kate}

\affiliation{European Organization for Nuclear Research (CERN), CH-1211, Gen\`eve 23, Switzerland}


\begin{abstract}
Axions are hypothetical particles that were postulated to solve one of the puzzles arising in the standard model of particle physics, namely the strong CP (Charge conjugation and Parity) problem. The new International AXion Observatory (IAXO) will incorporate the most promising solar axions detector to date, which is designed to enhance the sensitivity to the axion-photon coupling by one order of magnitude beyond the limits of the current state-of-the-art detector, the CERN Axion Solar Telescope (CAST). The IAXO detector relies on a high-magnetic field distributed over a very large volume to convert solar axions into X-ray photons. Inspired by the successful realization of the ATLAS barrel and end-cap toroids, a very large superconducting toroid is currently designed at CERN to provide the required magnetic field. This toroid will comprise eight, one meter wide and twenty one meter long, racetrack coils. The system is sized 5.2~m in diameter and 25~m in length. Its peak magnetic field is 5.4~T with a stored energy of 500~MJ. The magnetic field optimization process to arrive at maximum detector yield is described. In addition, materials selection and their structure and sizing has been determined by force and stress calculations. Thermal loads are estimated to size the necessary cryogenic power and the concept of a forced flow supercritical helium based cryogenic system is given. A quench simulation confirmed the quench protection scheme. 
\end{abstract}

\maketitle


\section{Introduction}

The mathematical description of the strong nuclear force includes a CP (Charge conjugation and Parity) symmetry violating term. However, there is no experimental evidence that quantum chromodynamics (QCD) breaks this symmetry. This "fine-tuning" problem is known as the strong CP problem. An appealing solution to this problem invokes an additional $U(1)$ chiral symmetry, the so-called Peccei-Quinn (PQ) symmetry \cite{Peccei:1977hh, Peccei:1977ur}. The spontaneous breaking of the PQ symmetry is associated with a light neutral pseudoscalar particle, the axion, which is closely related to the neutral pion \cite{Weinberg:1977ma, Wilczek:1977pj}. Axions are currently one of the most interesting non-baryonic candidates for dark matter in the universe. 

\begin{figure}[!b]
\centering
    \includegraphics[scale=0.5] {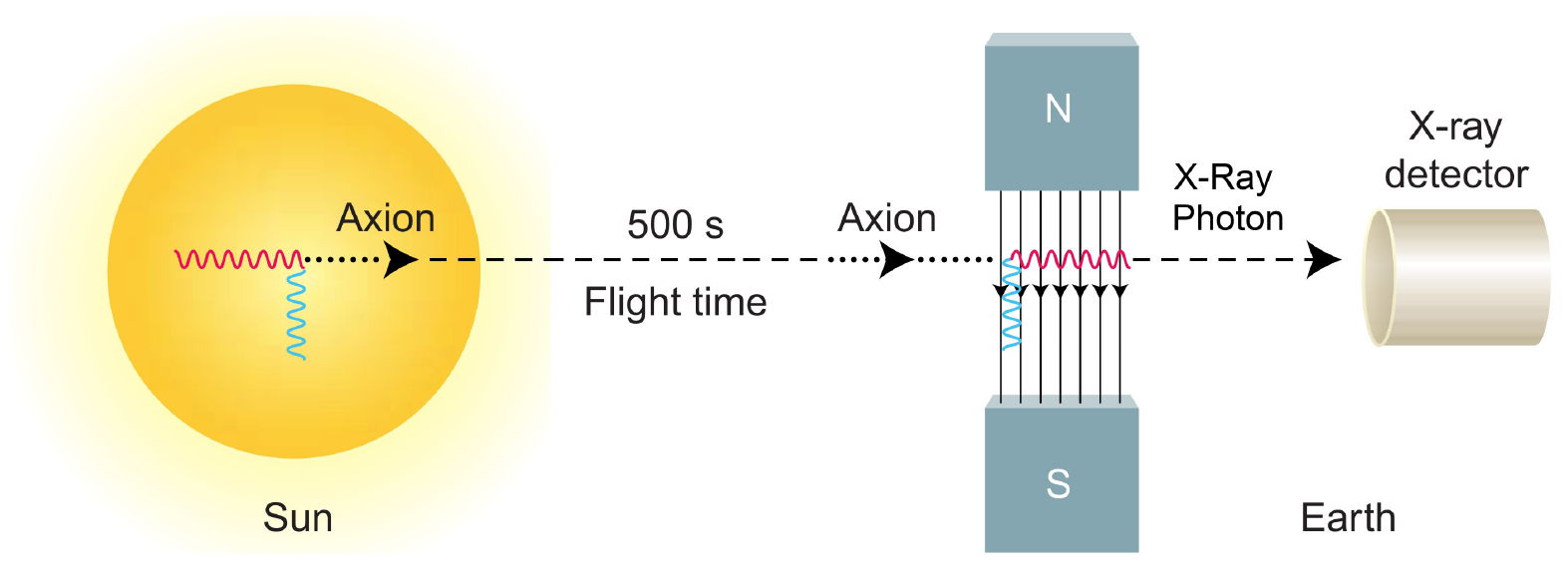}
    \caption{Schematic view of the solar axions detection concept: solar axions are predicted to convert to detectable X-ray photons when interacting with a magnetic field. Source: Science magazine Vol. 308 (2005).}
    \label{fig:1}
\end{figure}



\renewcommand{\arraystretch}{1.1}
\begin{table}[!ht]
\caption{Main design parameters of the IAXO toroidal magnet.}
\begin{center}
\begin{tabular*}{0.62\textwidth}{@{\extracolsep{\fill}} p{8 cm}  c } \hline
\textit{Property} & \textit{Value}\\ 
\hline

\textbf{Cryostat dimensions:} \hfill Overall length (m) & 25  \\
\hfill Outer diameter (m) & 5.2 \\
\hfill Cryostat volume (m$^3$) & $\sim$ 530 \\

\textbf{Toroid size:} \hfill Inner radius, $R_{in}$ (m)  & 1.0 \\
\hfill Outer radius, $R_{out}$ (m) & 2.0 \\
\hfill Inner axial length (m) & 21.0 \\
\hfill Outer axial length (m) & 21.8 \\

\textbf{Mass:} \hfill Conductor (tons) & 65 \\
\hfill Cold Mass (tons) & 130 \\
\hfill Cryostat (tons) & 35 \\
\hfill Total assembly (tons)  & $\sim$ 250 \\

\textbf{Coils:} \hfill Number of racetrack coils& 8 \\
\hfill Winding pack width (mm) & 384 \\
\hfill Winding pack height (mm) & 144 \\

\hfill Turns/coil & 180 \\
\hfill Nominal current, $I_{op}$ (kA) & 12 \\
\hfill Stored energy, $E$  (MJ) & 500 \\
\hfill Inductance (H) & 6.9 \\
\hfill Peak magnetic field, $B_p$ (T) & 5.4 \\
\hfill Average field in the bores (T) & 2.5  \\

\textbf{Conductor:} \hfill Overall size (mm$^2$) & 35 $\times$ 8 \\
\hfill Number of strands & 40 \\
\hfill Strand diameter (mm) & 1.3 \\
\hfill Critical current @ 5 T, $I_c$ (kA) & 58 \\
\hfill Operating temperature, $T_{op}$ (K) & 4.5 \\
\hfill Operational margin & 40\% \\
\hfill Temperature margin @ 5.4 T  (K) & 1.9 \\

\textbf{Heat Load:} \hfill  at 4.5 K (W) & $\sim$150 \\
\hfill at 60-80 K  (kW) & $\sim$1.6 \\
\hline \end{tabular*} 
\end{center}
\label{table1}
\end{table}

Axions interact very weakly with ordinary matter, thus making a direct observation practically impossible. However, axions are predicted to interact with magnetic fields in a way that allows them to convert to and from photons when passing through an area of high magnetic field. This property is used for axions searches in terrestrial experiments by means of photon detection (see Fig. \ref{fig:1}). In this work we define a new superconducting toroidal detector magnet that will be specifically designed for solar axions detection and serve as the fundamental part of IAXO, the International AXion Observatory. The IAXO project entails an immense upgrade of axion detection experiments, compared to the current state-of-the-art which is represented by the CAST experiment at CERN and using a 9~T, 9~m long, LHC dipole prototype magnet with twin $15$~cm$^2$ bores. IAXO features a dramatic enhancement of the present limits on axions search.

\section{Conceptual Design}

\subsection{Figure of Merit and Lay-Out Optimization}

The initial step towards a conceptual design of the IAXO magnet is to optimize its toroidal geometry so as to satisfy the primary design criterion: achieving a sensitivity to the coupling between axions and photons of one order of magnitude beyond the limit of the current experiment (CAST). To represent the magnet's merit within the experiment, the magnet's figure of merit (MFOM) is defined as $f_M = L^2B^2A$ \cite{IAXO}, where $L$ is the magnet length, $B$ the effective magnetic field and $A$ the aperture covered by the X-ray optics. Currently, CAST's MFOM is 21~T$^2$m$^4$. Accordingly, the IAXO magnet has to achieve an MFOM of 300 relative to CAST. Notice that the complete figure of merit of the experiment includes the tracking, detectors and optics parameters as well and is given in \cite{IAXO}.

When the magnet straight section length $L$ is set to 20 m, the MFOM is determined by the integral $\int B^2(x,y)dxdy$. The integration is performed solely across the \textit{open} area covered by the X-ray optics. Hence, the telescopes positioning must be determined to perform the integration. Upon minimizing the radial component with respect to the inner radius of the racetrack coil windings $R_{in}$, the optimized angular alignment of the telescopes is determined by the result of the integration. The latter shows that the MFOM is affected mostly by the fraction of the aperture of the telescopes exposed to X-rays, implying that the telescopes optimized alignment is in between each pair of coils to allow for maximum telescopes exposure. The complete lay-out optimization study, including references to other magnet designs that were considered during the geometrical study, is described in detail in \cite{asc}.

\begin{figure}[!t]
\centering
    \includegraphics[scale=0.5] {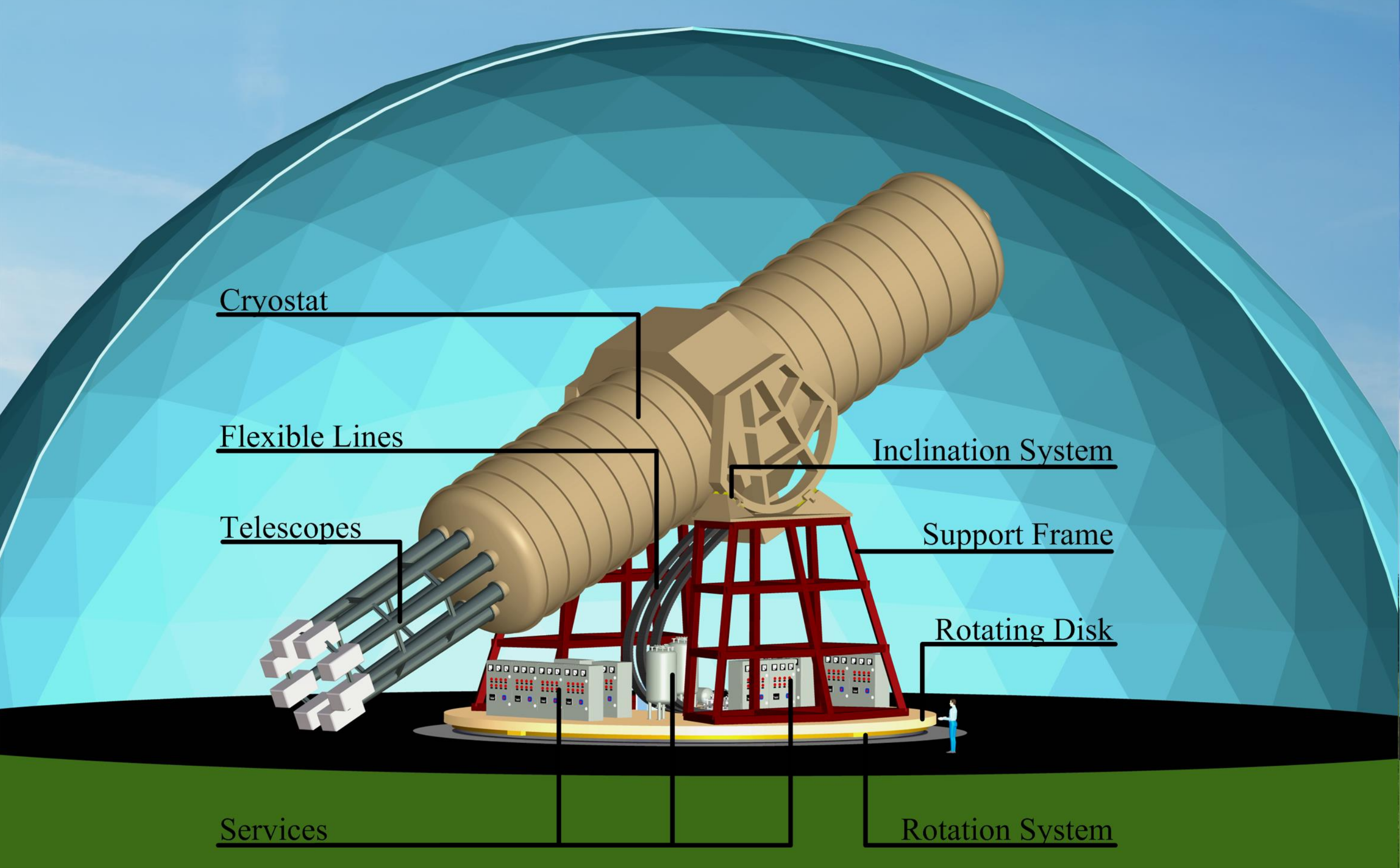}
    \caption{Schematic view of IAXO. Shown are the cryostat, eight telescopes, the flexible lines guiding services into the magnet, cryogenics and powering services units, inclination system and the rotating disk for horizontal movement. The dimensions of the system can be appreciated by a comparison to the human figure positioned by the rotating table.}
    \label{fig:2}
\end{figure}

Following the geometrical optimization study result, the design, presented in Fig. \ref{fig:2}, addresses all the experimental requirements of the magnet while relying on known and mostly well-proven engineering solutions, many of which were used in and developed for the ATLAS toroids engineered by CERN, INFN Milano and CEA Saclay. This ensures that the magnet is technically feasible to manufacture. The main properties of the toroid are listed in Table 1. The design essentially features a separation of the magnet system from the optical detection systems, which considerably simplifies the overall system integration. This also allows for eight open bores, which are centered and aligned in between the racetrack coils, in accordance with the geometrical study. The warm bores will simplify the fluent use of physics experimental instrumentation and the periodic maintenance of the system. 

The toroidal magnet consists of eight coils and their casing, an inner cylindrical support for the magnetic forces, keystone elements to support gravitational and magnetic loads, a thermal shield, a vacuum vessel and a movement system (see Figs. \ref{fig:2} to \ref{cs} and Table 1). Its mass is $\sim$250~tons with a stored energy of $\sim$500~MJ. The design criteria for the structural design study are defined as: a maximum deflection of 5 mm, a general stress limit of 50~MPa and a buckling factor of 5. The magnetic and structural designs are done using the ANSYS$^{\textregistered}$ 14.5 Workbench environment. The Maxwell 16.0 code is used to calculate 3D magnetic fields and Lorentz forces. The magnetic force load is linked to the static-structural branch to calculate stress and deformation. The eight bores are facing eight X-ray telescopes with a diameter of 600~mm and a focal length of $\sim$6~m. The diameter of the bores matches the diameter of the telescopes. The choice for an eight coils toroid with the given dimensions and eight 600~mm diameter telescopes and bores is also determined by a cost optimization within the anticipated budget for construction of the magnet of about 35 MCHF.  


\subsection{Conductor}

\begin{figure}
\centering
    \includegraphics[scale=0.26] {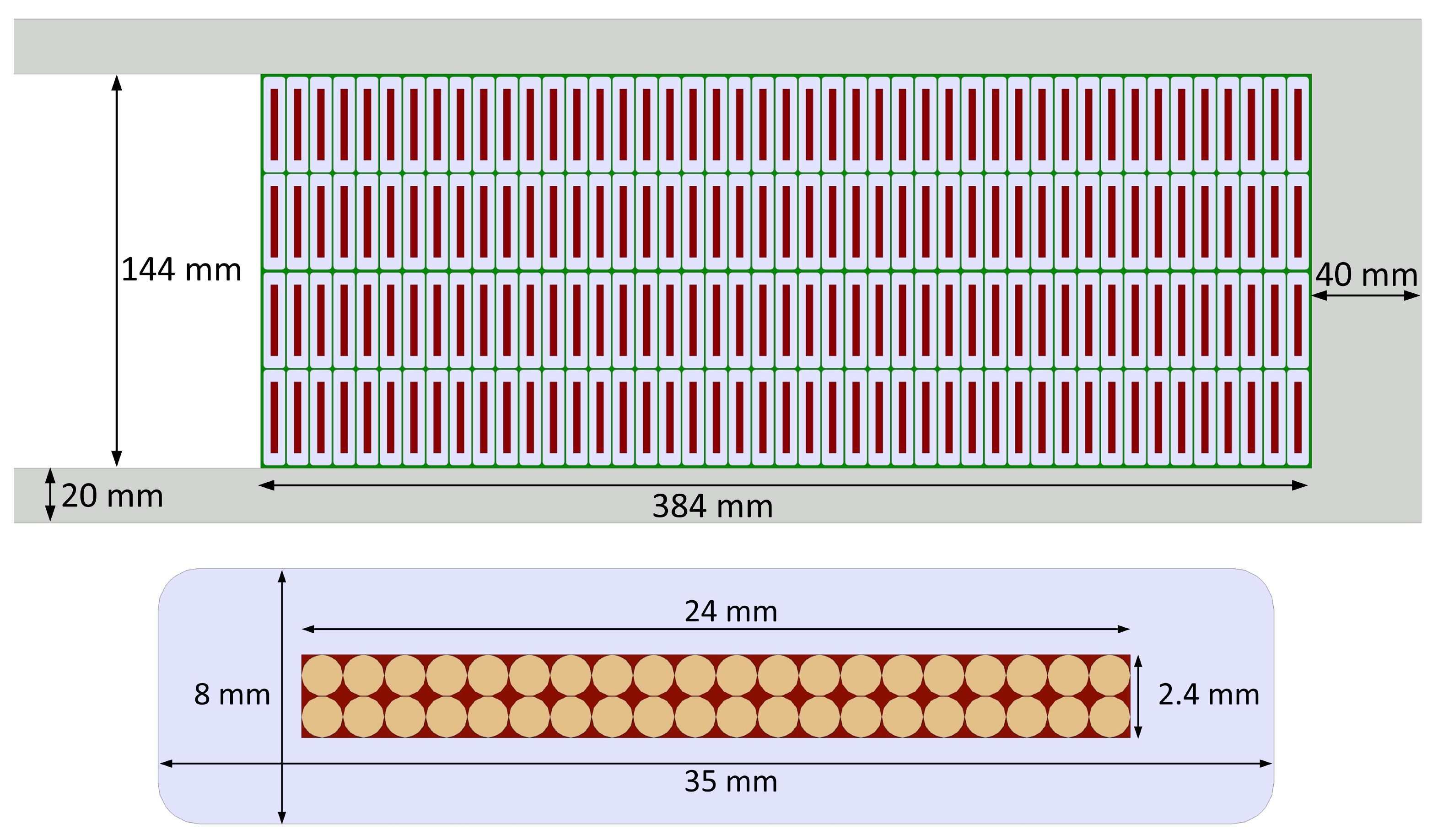}
    \caption{Cross section of the two double pancake winding packs, the coil casing (top) and the conductor with a 40 strands NbTi Rutherford cable embedded in a high purity Al stabilizer (bottom).}
    \label{coil}
\end{figure}

The conductor is shown in Fig. \ref{coil}. The large Rutherford type NbTi cable, comprising 40 strands of 1.3~mm diameter and a Cu/NbTi ratio of 1.1, is co-extruded within a high-purity and high RRR aluminum stabilizer, following the techniques used in the ATLAS and CMS detector magnets at CERN \cite{ATLAS}-\cite{CMS}. The use of a Rutherford cable as the superconducting element provides a high current density while maintaining high performance redundancy in the large number of strands. The Al stabilizer serves both magnet quench protection and stability of the superconductor. The conductor has a critical current of $I_c(5~\mbox{T},~ 4.5~\mbox{K}) =$ 58~kA.

\subsubsection{Peak Magnetic Field and Forces}

The peak magnetic field in the windings, which determines the operation point of the conductor and the temperature margin, is calculated at 5.4~T for a current of 12~kA per turn. Consequently, the net force acting on each racetrack coil is 19~MN, directed radially inwards. In order to minimize the forces acting on the bent sections, the racetrack coils are bent to a symmetric arc shape, with radius $R_{arc} = 0.5$~m.

\subsubsection{Stability Analysis}

In order to maximize its MFOM and extend the detection potential of the experiment, the IAXO toroid requires maintaining the highest possible magnetic field. Nevertheless, acceptable operational current and temperature margins are compulsory in order to ensure proper and safe operation. For a two double pancake configuration with 180 turns and engineering current density $J_{eng} = 40$~A/mm$^2$, the peak magnetic field is $B_p = 5.4$~T. The critical magnetic field corresponding to the magnet load line is 8.8~T at 65~A/mm$^2$. Hence, IAXO's magnet is expected to work at about 60\% on the load line, which sets the operational current margin to 40\%.

The temperature margin calculation is based on an operational temperature of $T_{op} = 4.5$~K and a peak magnetic field of $B_p = 5.4$~T. A coil with two double pancakes and 45 turns per pancake satisfies a temperature margin of 1.9~K, while yielding an MFOM of 300, relative to CAST, thus satisfying the principal design criterion.

\subsection{Quench Protection}

The adiabatic temperature rise due to a uniformly distributed quench is about 100~K. The quench protection is based on an active protection system with multiple quench heaters positioned along each of the eight racetrack coils to enforce a fast and reliable quench propagation and so a homogenous cold mass temperature after a quench. The concept of the quench protection system is to rely on a simple, robust and straight forward detection circuits, usage of simple electronics and at least a three fold redundancy in order to reduce failure probability. 

Correspondingly, the magnet power supply is connected at its DC outputs to two breakers, which open the electrical lines to the magnet. When a quench is detected and verified, the two breakers open to quickly separate the magnet from the power supply and a quench is initiated in all coils simultaneously by activating all the quench heaters. In addition, the current is discharged through a dump resistor with low resistance, connected in series to a diode unit. This discharge mode, so-called the fast dump mode, is characterized by an internal dump of the magnet's stored energy. Upon grounding the magnet, the fast discharge scheme ensures that the discharge voltage excitation is kept low and that the stored energy is uniformly dissipated in the windings.

Under normal operation, the toroid may be discharged through the dump-resistor in a passive run down mode (slow discharge mode). Slow discharge is also the safety dump mode activated in the case of a minor fault, as described below.

The longitudinal normal zone propagation velocity is 6~m/s. The velocity was calculated by using COMSOL 4.3a coupled multiphysics modules in a 2D adiabatic model. Therefore, the normal zone will propagate around an entire coil (43~m) in 3.5~sec, where usual quench protection system initiation delay times are $\sim$ 1~s.

\subsection{Cold Mass}

The cold mass operating temperature is 4.5~K and its mass is approximately 130~tons. The cold mass consist of eight coils, with two double pancakes per coil, which form the toroid geometry, and a central cylinder designed to support the magnetic force load. The coils are embedded in Al5083 alloy casings, which are attached to the support cylinder at their inner edge. The casings are designed to minimize coil deflection due to the magnetic forces. 

A coil, shown in Fig. \ref{coil}, comprises two double pancake windings separated by a 1 mm layer of insulation. The coils will be impregnated for final bonding and pre-stressed within their casing to minimize shear stress and prevent cracks and gaps appearing due to magnetic forces.

\begin{figure}[!t]
\centering
    \includegraphics[scale=0.47] {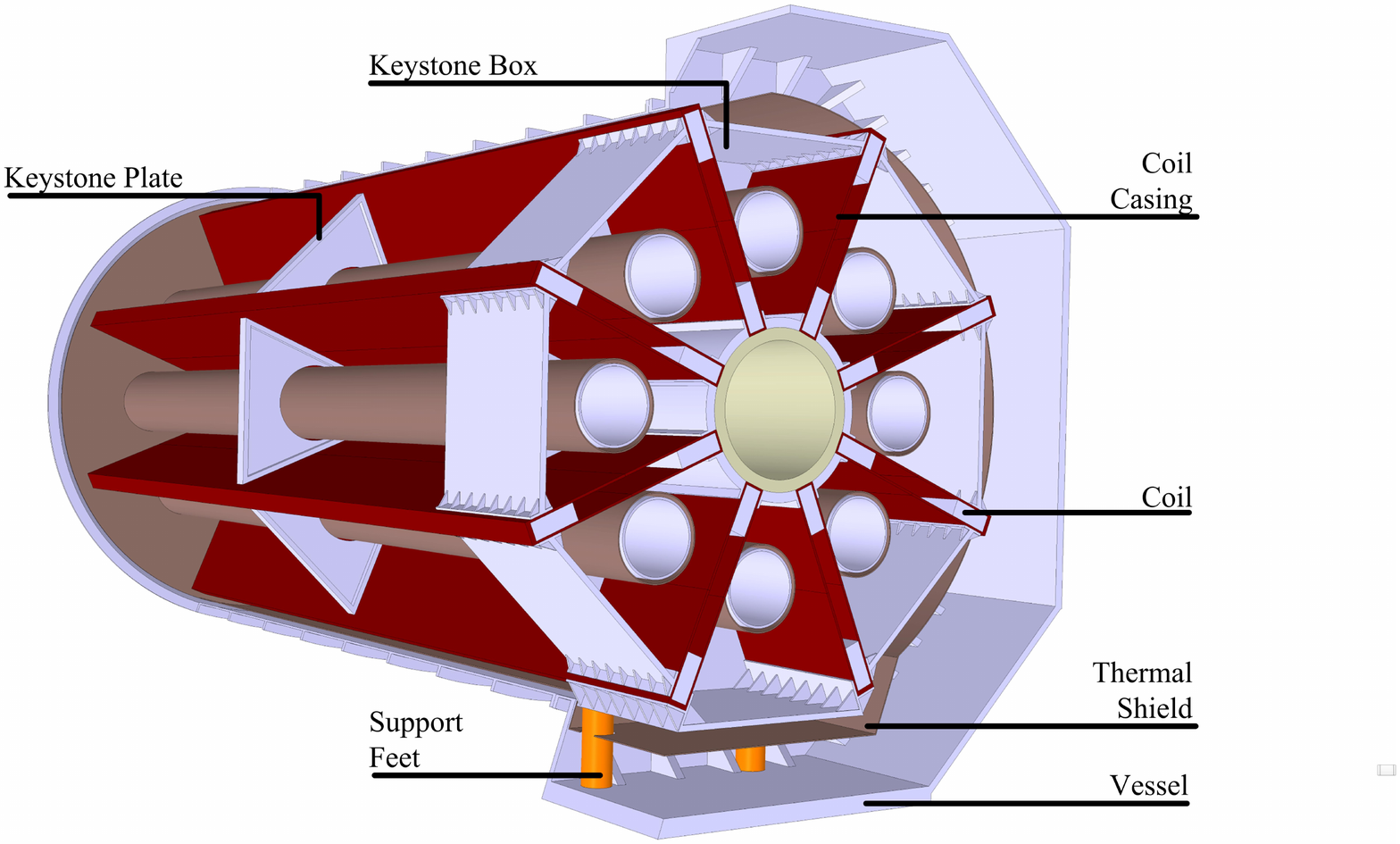}
    \caption{Mid-plane cut of the cryostat with an exposed cold mass, showing the cold mass and its supports, surrounded by a thermal shield, and the vacuum vessel. The open bores will simplify the use of scientific instrumentation.}
    \label{cs}
\end{figure}

To increase the stiffness of the cold mass structure and maintain the toroidal shape under gravitational and magnetic loads, and to support the warm bores, eight Al5083 keystone boxes and 16 keystone plates are connected in between each pair of coils, as shown in Fig. \ref{cs}. The keystone boxes are attached to the support cylinder at the center of mass of the entire system (i.e. including the telescopes and detectors) and the keystone plates are attached at half-length between the keystone boxes and the coils ends.

\subsection{Cryostat and Its Movement System}

The design of the cryostat is based on a rigid central part, placed at the center of mass of the entire system and serves as a fixed support point of the cold mass, with two large cylinders and two end plates enclosing it to seal the vacuum vessel. In addition, eight cylindrical open bores are placed in between the end plates. The vessel is optimized to sustain the atmospheric pressure difference and the gravitational load, while being as light and thin as possible. The Al5083 rigid central piece is 70~mm thick with a thicker 150~mm bottom plate to support the mass of the cold mass. Using two end flanges at the vessel's rims, as well as periodic reinforcement ribs at 1.35~m intervals along both cylinders, the structural requirements are met for a 20~mm thick Al5083 vessel with two 30~mm thick torispherical Kl\"opper shaped end plates. The 10~mm wall thickness of the eight cylindrical bores is minimized in order for the bores to be placed as close as possible to the racetracks coils inner radius, thereby maximizing the MFOM. 

The cold mass is fixed to the central part of the cryostat. The cold mass supports are made of four G10 feet, connecting the reinforced bottom keystone box (referred to as KSB8) to the central part of the cryostat and transfer the weight load of the cold mass to the cryostat. KSB8 also provides a thermal property: the cold mass supports are not directly attached to the coils casings, thereby reducing the heat load on the windings and affecting less the stability of the magnet. The support feet are thermally connected to the thermal shield with copper braids, further reducing the heat load on KSB8. Moreover, KSB8 can be directly cooled to ensure that the magnet stability margins remain at the desired level. 

Mechanical stops, which counteract forces along a specific axis, will be introduced at both ends of the cold mass to reduce the stress on the fixed support feet when the magnet is positioned at different inclination angles.

Searching for solar axions, the IAXO detectors need to track the sun for the longest possible period in order to increase the data-taking efficiency. Thus, the magnet needs to be rotated both horizontally and vertically by the largest possible angles. For vertical inclination a $\pm$ 25$^\circ$ movement is required, while the horizontal rotation should be stretched to a full 360$^\circ$ rotation before the magnet revolves back at a faster speed to its starting position. 

The 250 tons magnet system will be supported at the center of mass of  the entire system through the cryostat central part (see Fig. \ref{fig:2}), thus minimizing the torques acting on the support structure and allowing for simple rotation and inclination mechanisms. Accordingly, an altitude-over-azimuth mount configuration was chosen to support and rotate the magnet system. This mechanically simple mount, commonly used for very large telescopes, allows to separate vertical and horizontal rotations. The vertical movement is performed by two semi-circular structures (C-rings) with extension sections which are attached to the central part of the vacuum vessel. The C-rings distribute support forces from the rigid central part of the vessel to the C-rings pedestals, equipped with elevation hydrostatic pads and drives. The pedestals are mounted on top of a 6.5~m high structural steel support frame which is situated on a wide rotating structural steel disk. The rotation of the disk is generated by a set of roller drives on a circular rail system.

The required magnet services, providing vacuum, helium supply, current and controls, are placed on top of the disk to couple their position to the horizontal rotation of the magnet. The magnet services are connected via a turret aligned with the rotation axis, thus simplifying the flexible cables and transfer lines arrangement. A set of flexible chains are guiding the services lines and cables from the different services boxes to the stationary connection point.

\subsection{Cryogenics}

The coil windings are cooled by conduction at a temperature of 4.5~K. The conceptual design of the cryogenic system is based on cooling with a forced flow of sub-cooled liquid helium at supercritical pressure. This avoids two-phase flow, that would be very difficult to control due to the continuously changing inclination angle of the magnet, within the magnet cryostat. The coolant flows in a piping system attached to the coil casings, allowing for conduction cooling in a manner similar to the ATLAS toroids \cite{ATLAS, ATLAS2}. The decision for using this design is following the same philosophy of our concept: a known technology with a low-cost proven solution and most reliable.

The heat load on the magnet by radiation and conduction is $\sim$150~W at 4.5~K. In addition comes the thermal shield heat load of $\sim$1.6~kW. An acceptable thermal design goal is to limit the temperature rise in the coils to 0.1~K above the coolant temperature under the given heat loads. 

\begin{figure}[!b]
\centering
    \includegraphics[scale=0.42] {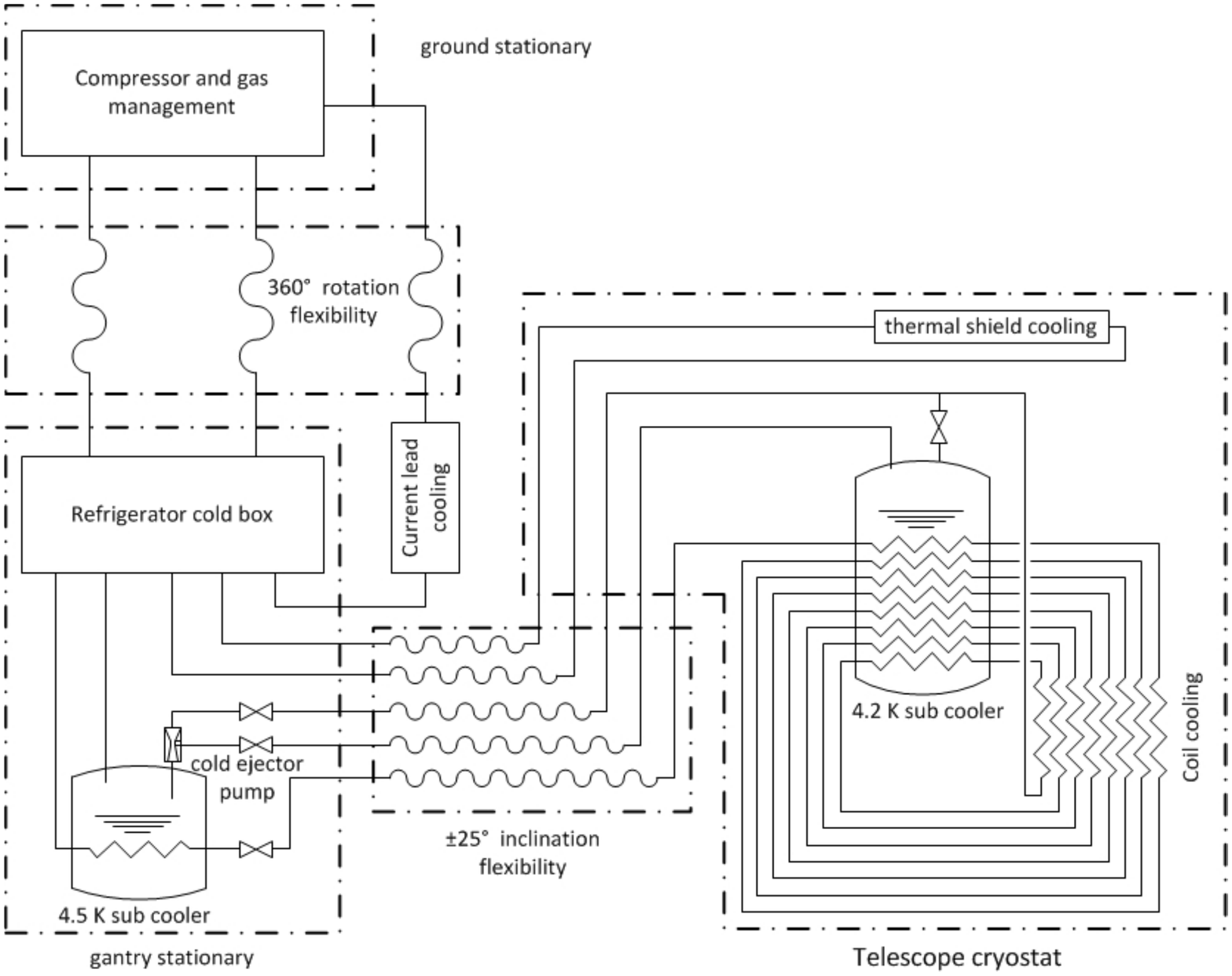}
    \caption{Flow diagram of the cryogenic system of the IAXO magnet.}
    \label{cryo}   
\end{figure}

Fig. \ref{cryo} shows a schematic flow diagram of the cryogenic system concept. It features the helium compression and gas management that is ground stationary. The refrigerator cold box, current leads cryostat and a 4.5~K helium bath are integrated on the rotating disk that carries the structure of the helioscope. A helium bath operating at 4.2~K is connected to the magnet cryostat to follow its movement.

The magnet coils are cooled by a helium flow of 23~g/s, supplied at about 300~kPa and 4.6~K and sub-cooled in the 4.5~K bath. Before entering the cooling circuit of the first coil, the flow is cooled to 4.3~K in the 4.2~K bath. After passing through the cooling channels of each coil, the helium, then at below 4.5~K, is re-cooled in the sub-cooler of the magnet cryostat. As the flow returns from the last magnet coil channels, part of the helium is used to supply the 4.2~K sub-cooler and the remaining gas supplies the 4.5~K bath on the rotating disk. The latter flow is also used as a drive flow for a cold injector pump that pumps the 4.2~K sub-cooler at ambient pressure. 

The thermal shield is cooled by a flow of 16~g/s gas at 16~bar between 40~K and 80~K. The cooling of the current leads is supplied at 20~K and 1.2~bar, which corresponds to the cooling used for the HTS current leads of the LHC machine \cite{cryo1}. The path of the superconducting cables to the magnets is not shown in Fig. \ref{cryo} but they could for example be integrated in the helium supply line.

The total equivalent capacity of the refrigerator results in a 360~W isothermal load at 4.5~K. Thus, the refrigerator cold box will be compact enough to be integrated together with the cryostat of the current leads on the gantry that is rotating with the helioscope. All cryogenic lines between the refrigerator and the magnet cryostat will therefore only need to compensate for the $\pm$ 25$^\circ$ inclination, and not for the 360$^\circ$ rotation that will be followed only by ambient temperature lines.

\section{Fault Scenarios and System Reliability}

The IAXO magnet system is a complex combination of subsystems which work in harmony. Therefore, the anticipation of fault scenarios and the basic operational strategy in case of such failures should be dealt with already at the design stage. Here, we identify and describe the major fault cases which could interrupt the normal operation of the system: 

\begin{itemize}

\item \textit{Cryogen leak:} Minor leaks in the cryogenics pipes will result in exceeding the vacuum system trip limits. In this case the safety system will initiate a slow magnet discharge. In the case of a rupture in the cryogen lines a rapid pressure rise in the vessel will occur. The vessel will remain protected by a set of relief valves, while a fast shut down of the system will be initiated. 

\item \textit{Vacuum failure:} The vacuum system is supported by safety valves, thus decreasing the probability for a catastrophic vacuum failure considerably. Normal vacuum system faults will be dealt with by hard wired interlocks.

\item \textit{Quench protection system failure:} Total failure of the quench detection system or the heaters system will be avoided by using multiple detectors and heaters to give redundancy to the system. Nonetheless, the coils and conductors are designed to sustain even such fault conditions so that a complete quench system failure will not lead to coil nor conductor damage.

\item \textit{Power failure:} If the mains power will fail to supply power to the magnet control systems, the power supply will be maintained by a UPS. Nonetheless, such a fault scenario will initiate a slow discharge of the magnet.

\item \textit{Refrigeration supply failure:} A failure to supply cooling power from the refrigerator cold box to the current leads and bus-bars, thermal shield or the 4.2~K sub-cooler will initiate a slow discharge process.

\item \textit{Water supply failure:} Water supply failure to the power converter, vacuum pumps, etc. will result in a slow discharge.

\item \textit{Air supply failure:} Air supply is required for the ongoing operation of the vacuum system and the cryogenic system. A deficient air supply to these systems will lead to a slow discharge of the magnet.

\item \textit{Seismic disturbances:} The structure and movement system must sustain an additional sidewards load of at least 1.2g, which may be caused by a moderate seismic activity.

\end{itemize}

System reliability is an important issue when designing a complex assembly of subsystems such as the IAXO magnet system, let alone when the system is required to operate for long periods of time without exterior interference. Some key factors are to be noted when defining the magnet system's reliability: A fast discharge of the magnet should be initiated only when the magnet, experiment as a whole or personal safety is in danger. In all other cases a slow energy dump should take place. A UPS unit will maintain key services in order to enable a safe and controlled slow discharge in extreme cases. Lastly, routine maintenance is essential to avoid false magnet discharges.

\section{Assembly Procedure and Integration}

The IAXO detector will be placed in a light and confined structure, such as a dome or a framed tent, which will serve as the main site for the experiment. For this reason the assembly requires a hall with enough space to allow for the large tooling and infrastructure needed for the final cold mass and cryostat integration. 

The assembly of the cold mass and the cryostat will be performed in five main steps: First, each of the eight warm bores, surrounded by 30 layers of super-insulation and a thermal shield, will be connected to one keystone box and two keystone plates to form eight sub-units. These sub-units will be attached, together with the coils casings, to the cold mass central support cylinder in order to assemble the complete cold mass. Cooling circuits will be installed and bonded to the surface of the coils casings already during fabrication. Additional cooling pipes will be attached to the cold mass when the latter is assembled. Next, the complete cold mass will be connected to the central part of the cryostat, where the cylindrical cold mass G10 based supports will be inserted into their their designated slots. The two cylindrical parts of the vessel will be connected to the central part, followed by the enclosure of the cryostat by the two Kl\"opper end plates which will be connected to the end flanges on both sides of the cryostat cylinders and to the bores. Lastly, the magnet vessel will be transferred to the main site where it will be attached to the movement system. The installation of services lines to the services turret, as well as the integration of the magnet system with the rest of the experiment's systems, will be performed at the last stage of system integration in the main site.

\section{Conclusion}

 The design of the new IAXO toroidal superconducting magnet satisfies the design criterion of increasing the solar axion searches sensitivity by at least one order of magnitude. The design relies on known engineering solutions and manufacturing techniques and thus is technically feasible to manufacture. With the inclusion of eight open warm bores, the design also allows maximum flexibility for the experimentalists. This serves not only to maximize and extend the detection potential of axions searches, but also of more exotic particles, commonly known as Axion Like Particles (ALPs). The magnet system and optical detectors are separated, thus allowing for a parallel effort of development, construction and initial operation, thereby also minimizing cost and risks. In the coming years the design will be further detailed and project funding will be proposed.

\end{document}